# Doping icosahedral $Fe_{13}$ with 3d transition elements

Anteneh G Tefera and Mogus D Mochena

Department of Physics, Florida A & M University, Tallahassee, FL 32307

## Abstract

Our density functional theory calculations for $Fe_{13-n}M_n$ for M = Sc, Ti, V, Cr, Mn, Co, Ni, and Cu up to n = 4 show that the icosahedral symmetry of $Fe_{13}$, albeit minor changes in bond lengths, is robust despite doping and is retained for all homotops of Mn, Co, Ni and Cu. Based on analysis of density of states of the doped cluster, adsorption of carbon atom, and adhesion energies for fragments of single walled carbon nanotubes, we propose a core-shell type structure with a central Mn atom surrounded by Fe surface shell atoms as the most favorable doped nanocatalyst for SWCNT nucleation and growth subject to constraints of retention of icosahedral symmetry by the doped cluster. For doping beyond the central atom involving the surface shell of the icosahedron, Ni is the best candidate.

## 1. Introduction

Monoatomic and polyatomic nanoclusters of transition metals (TM) have been the subject of intense research interest for the last two decades.[1] Recently bimetallic TM nanocatalysts have attracted more attention as prospective catalysts that could lead to controlled synthesis of single walled carbon nanotubes (SWCNTs).[2] Chirality controlled synthesis of SWCNTs has been elusive for two decades, but recently it has been reported that by tuning the composition of bimetallic catalysts, chiral selectivity can be improved considerably. Most of the synthesis work has involved doping the most widely used monometallic TM catalysts Fe, Co and Ni, for example, FeMo, FeRu, FeCu, CoMo, CoMn, CoCr, CoW, CoPd, CoTi[3-21], NiFe.[2] It seems the choice of the doping element is random, and the underlying mechanism behind improved selectivity is not well understood.[22] A systematic approach is, therefore, necessary to understand the catalytic process for a rational design of the catalysts to enhance selectivity.

The Fe clusters are the archetypical catalysts widely used to synthesize SWCNTs ever since their discovery. $Fe_{13}$ with a diameter of approximately 0.47 nm is on the lower end of the catalyst size spectrum for SWCNT synthesis. Loebick et al reported growing subnanometer SWCNTs of 0.5 to 0.7 nm diameter at 600 ºC, 700 ºC and 800 ºC using CoMn bimetallic catalysts supported on MCM-41 silica templates.[23] On computational side, Raty et al studied early stage growth of SWCNT on $Fe_{55}$ with a diameter of ~ 1 nm, but also simulated using 13 - atom metallic clusters of Fe and Au to compute binding energies of the clusters to carbonaceous fragments of $C_{13}H_9$ and $C_{10}H_8$.[24] Feng et al modeled with $M_{13}$ icosahedral clusters (M = Fe, Co, Ni, Cu, Pd and Au) to determine adhesion energies of (3, 3) and (5, 0) SWCNT fragments to the metal clusters in addition to $M_{55}$.[25] We have recently looked into possibility of chirally selective growth promoted by the icosahedral symmetry of $Fe_{13}$ at low temperature.[26, 27] The retention of the symmetry is important in our model since we argue the symmetry of subnanometer catalyst can lead to chirally selective growth of the SWCNT at low temperature. Cantaro et al reported synthesizing SWCNT at 350 ºC.[28] Others have claimed that even lower temperatures with plasma enhanced CVD can lead to SWCNT growth.[29] The essence of this paper, however, is not demonstrating growth from doped $Fe_{13}$, but rather understanding the effect of doping on the icosahedral symmetry and electronic properties of $Fe_{13}$ and find a mechanism for selecting the dopant subject to certain constraints. $Fe_{13}$ is chosen rather than the more realistic $Fe_{55}$ because the configurational permutations (homotops) of the dopants in $Fe_{55}$ would be too many and costly to handle computationally. Nevertheless we assume $Fe_{13}$ would provide us with insight into the doping effects without complicating the physics with many possibilities. In this work, we consider the ground sate configuration of $Fe_{13}$ as a reference configuration and assume the magnetic correlations are set in it. This assumption reduces the number of homotops further since the dopants retain the same magnetic moment orientation as the Fe atoms they replace. Our main goal in this work is to find out if the icosahedral symmetry can be retained despite doping and if the surface states introduced as a result could affect the adsorption of carbon and adhesion of SWCNT fragments. Our simplified assumption for the magnetic correlation could result in loss of some surface states, but we think the work will capture the general trend of the effect of doping.

The rest of the paper is organized as follows. In section 2 we present the computational method employed, followed by results and discussion in section 3, and conclusion in section 4.

## 2. Computational Details

The low-energy structures and electronic properties of $Fe_{13-n}M_n$ (n = 1-4, M = Sc, Ti, V, Cr, Mn, Co, Ni, Cu) were computed with density functional theory (DFT) as implemented in Vienna Ab Initio Simulation Package (VASP).[30-32] The 3d and 4s electrons were treated as valence electrons for the transition-metal elements. The computations were performed with the projector augmented wave (PAW) potentials with a plane - wave energy cut off of 300 eV for $Fe_{13-n}M_n$ cluster and 420 eV for $Fe_{13-n}M_nC_p$ clusters for p=1 to calculate the adsorption energy of a carbon atom and p = 40 to calculate the adhesion energy of a SWCNT fragment. The reciprocal space integrations were carried out at gamma point. A vacuum region of 12 Å surrounded the cluster to avoid the interaction between the periodic cluster images. A linear mixing of input and output charge densities in the Pulay scheme was applied during the self-consistency loop. To calibrate the VASP input parameters, the ground state structure of $Fe_{13}$ cluster was determined first with all electron DFT in the generalized gradient approximation (GGA) using Perdew-Wang exchange – correlation functional[33] as implemented in GAUSSIAN '03 code.[34] The resulting ground state structure from GAUSSIAN was reproduced with VASP using the exchange correlation function of Perdew and Wang with Vosko Wilk and Nusair interpolation technique[35] for correlation part of the exchange correlation functional in spin-polarized GGA. A conjugate gradient algorithm was used to optimize the symmetry - unrestricted geometry at intervals of 0.5 fs until convergence is attained when all the forces are less than the threshold value of 0.001 eV/Å.

The optimized structures were then checked for their stability as the energetic stability of binary nanoclusters is an important index in determining their magic sizes and compositions. We determined the stability using the second energy difference, $\Delta E_2$, defined as[36]

$$\Delta_2 E(A_m B_n) = E(A_{m+1} B_{n-1}) + E(A_{m-1} B_{n+1}) - 2 E(A_m B_n) \quad (1)$$

Next we determined which one of the stable structures for the different dopants would be preferred in the synthesis of SWCNTs by comparing the adhesion energies of SWCNT fragments to the nanocatalysts. Fragments of SWCNT, passivated at one end with hydrogen atoms, were attached to the stable, optimized structures. The adhesion energies were calculated as

$$\Delta_{adhesion} = E(nanocluster+SWCNT+H) - E(nanocluster) - E(SWCNT+H\ atoms) \quad (2)$$

### 3. Results and Discussion

#### 3.1 Structure

We showed in refs. 26 and 27 how the icosahedral symmetry of $Fe_{13}$ leads to (5, 0) SWCNT. $Fe_{13}$ has an icosahedral geometry with 20 triangular faces, 12 vertices that are symmetrically equivalent and one atom at the center. In order to handle the large number of homotops[36], we started off our geometry optimization of $Fe_{13-n}M_n$ for n = 1 - 4 with $Fe_{12}M$ and increased the number of dopants, building successive structures based on results of previous ones. For $Fe_{12}M$, we replaced one of $Fe_{13}$ atoms with M on two different, possible sites, one at the center and the other on one of the vertices, and optimized the structures. We label these structures homotop one and homotop two. As expected, homotop one had lower energies for all the TM elements. The icosahedral geometry remained intact for a substitution at the center since the Fe atoms surround M symmetrically.

Next, we considered $Fe_{11}M_2$. Based on $Fe_{12}M$ results, the first M is placed at the center. The second M can then be placed on any of the vertices, which are equidistant from the center. Since this is the only possibility, we proceeded to $Fe_{10}M_3$.

For a small sized nanocluster of $Fe_{13}$, doping with three atoms comprises 23% of the whole structure. Among the possible homotops of $Fe_{10}M_3$ having an M atom at the center and the other two on the surface of the cluster resulting from substitution, we optimized the seven possible icosahedral structures as shown in Fig. 1 on sites 1-7. The orientation of the nanocluster is kept the same for all so as to keep track of the sites of substitution. The

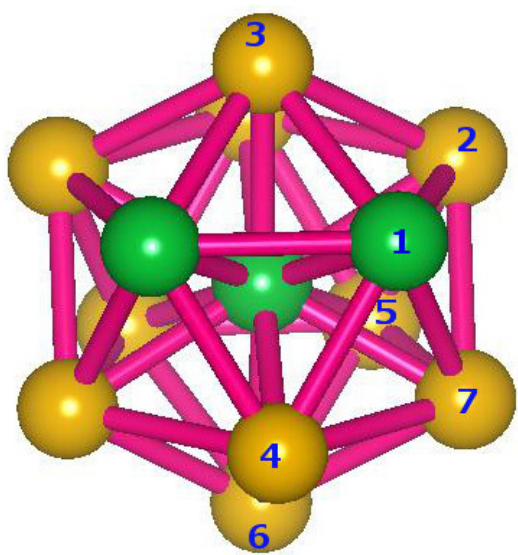

FIG. 1. $Fe_{13-n}M_n$ for n =3. The gold atoms are Fe and the green ones are M. The two unlabeled M atoms are fixed, and the third atom is substituted in one of the possible labeled sites resulting in seven homotops.

structure therefore has two apexes on both ends and two pentagonal rings in between the apexes. The other four remaining sites are symmetrically equivalent to one of the chosen sites, therefore, are left out. The results of optimization varied according to the size of the elements. The lighter ones, Sc and Ti, resulted in distortion from icosahedral symmetry, with bond breaking between the neighboring M atoms that opened up the structure. So Sc and Ti were ruled out. The lowest energy structure of V among the seven possible homotops is distorted from icosahedral geometry; therefore, we ruled out V as well. The rest of the 3d TM elements resulted in slight distortions, with minor changes in bond lengths, of the icosahedron, including the homotops with higher energies. The lowest-energy structures and total magnetic moments for M = Cr, Mn, Co, Ni, and Cu are given in Fig. 2. Our magnetic moment of the parent $Fe_{13}$ nanocluster is 34 $\mu_B$ which agrees well

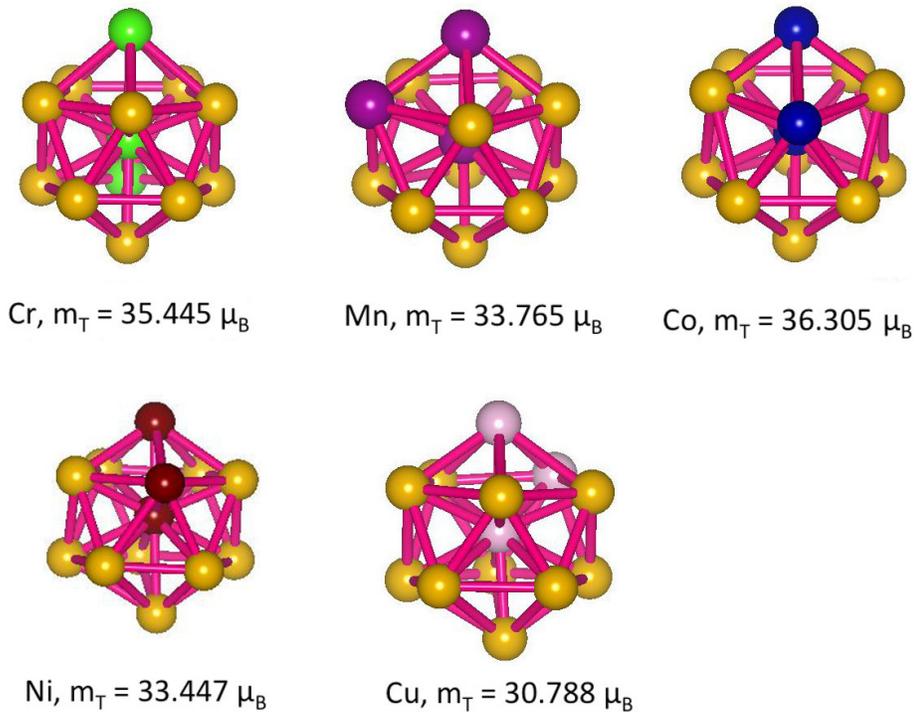

Cr, $m_T$ = 35.445 $\mu_B$    Mn, $m_T$ = 33.765 $\mu_B$    Co, $m_T$ = 36.305 $\mu_B$

Ni, $m_T$ = 33.447 $\mu_B$    Cu, $m_T$ = 30.788 $\mu_B$

FIG. 2. $Fe_{13-n}M_n$ for n = 3. The gold atoms are Fe and the other colors are M. The structures shown are the lowest energy homotops for M = Cr, Mn, Co, Ni, and Cu where $m_T$ is the total magnetic moment.

with experimental values of 2.6 ± 0.4 $\mu_B$ per atom.[37] The doping affects the magnetic moments by different amounts for the different dopants. Next we optimized $Fe_9M_4$, for which the doping comprises almost 31% of the whole structure. The input structures were constructed on the basis of the lowest energy structures of $Fe_{10}M_3$. We optimized seven, icosahedral homotops. The remaining three sites are symmetrically equivalent to one of the chosen sites and are left out. The resulting optimized lowest energy structures retain their icosahedral symmetry as shown in Fig. 3. All the remaining six structures with higher energies retain their icosahedral symmetry as well for each element except chromium. For Cr, one of the structures is distorted and is 60 meV above the lowest energy. This shows that icosahedral symmetry of the low-energy structures is quite robust and is not affected by the doping in general.

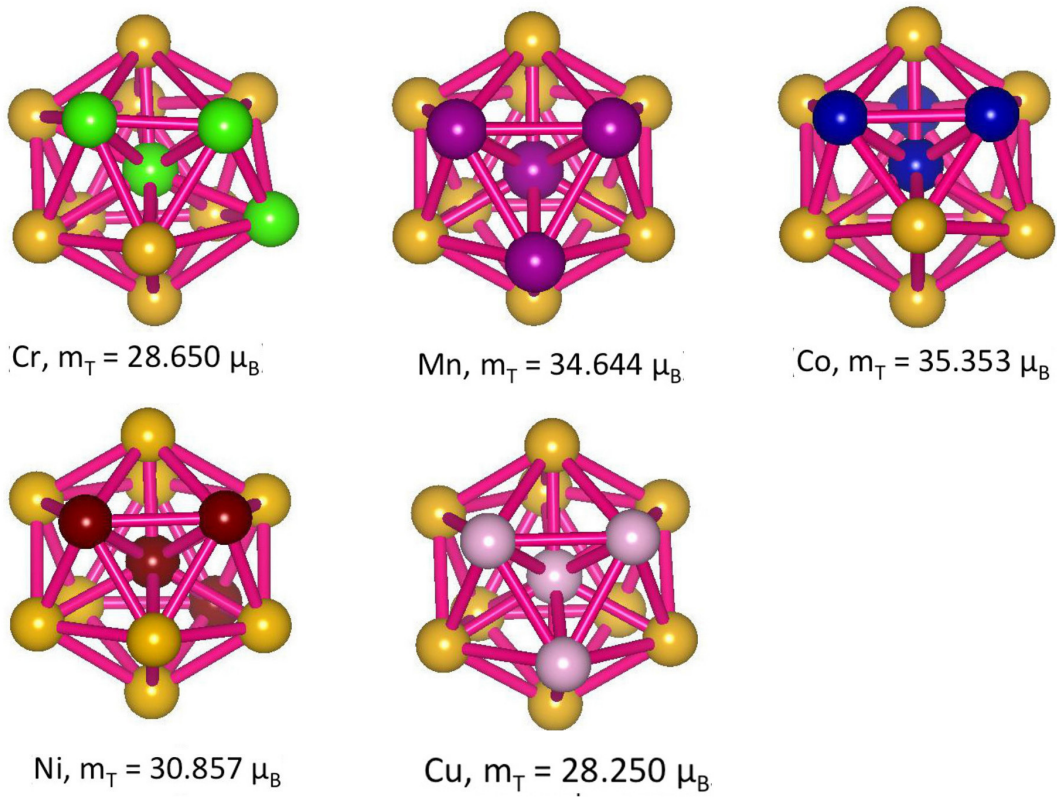

Cr, $m_T$ = 28.650 $\mu_B$

Mn, $m_T$ = 34.644 $\mu_B$

Co, $m_T$ = 35.353 $\mu_B$

Ni, $m_T$ = 30.857 $\mu_B$

Cu, $m_T$ = 28.250 $\mu_B$

FIG. 3. $Fe_{13-n}M_n$ for n = 4. The gold atoms are Fe and the other colors are M. The structures shown are the lowest energy homotops for M = Cr, Mn, Co, Ni, and Cu, where $m_T$ is the total magnetic moment.

### 3.2 Stability

One way of measuring the stability of a bimetallic cluster is to compare the stability of the doped cluster to adjacent compositions using the second difference in energy, $\Delta_2 E$, as given in equation (1). For pure clusters, $\Delta_2 E$ usually compares sizes differing by one atom, whereas for binary clusters, it is defined for fixed size and variable composition. Clusters with high relative stability correspond to peaks in $\Delta_2 E$ as function of concentration of dopants. In Fig. 4, $\Delta_2 E$ is plotted as function of the concentration for $Fe_{13-n}M_n$, n = 1-3 and M = Cr, Mn, Co, Ni and Cu. The results clearly show that the structures for n = 1 are the most stable and are consistent with similar work as reported in ref. 36. This stability could be of relevance for a large nanocluster with core/shell geometry, suggesting the core region could be substituted with desired atoms without affecting the overall icosahedral geometry, for instance in $Fe_{55}$.

Despite the maximum stability for n=1, it is relevant to look into the effect of doping with increased n. Surface geometry and surface states are important for SWCNT as carbon species nucleate and grow. Therefore, we increased the doping up to n = 4 and the results of $\Delta_2 E$ are given in Fig. 4. The dotted horizontal line represents the $\Delta_2 E$ for pure $Fe_{13}$ cluster. The plots for Cr, Co and Ni are above the threshold level of $\Delta E_2 = 0$ up to n = 3 and drop. They barely touch the threshold at n = 2. The Mn curve decreases below threshold slightly from n = 2 to n = 3 and then gradually continues to rise. The plot for Cu is quite distinct from the rest. It drops significantly from n = 1 to n = 2 and then continues to rise.

To study the effects of one element on the other, it is obvious one can only dope with less than 50 % of the dopant. We have doped up to n = 4, dopant concentration of 31 %, which is quite significant. Further doping to n = 5 and n = 6 increases the number of homotops significantly, so we stopped at n = 4. From the results we conclude doping $Fe_{13}$ with more than 23 % (i.e. n = 3) of Cr, Co and Ni results in less stable structure, and for Mn and Cu it is possible to dope with higher percentage and get a stable structure. We cannot explicitly state how high the percentage goes since we stopped at n = 4.

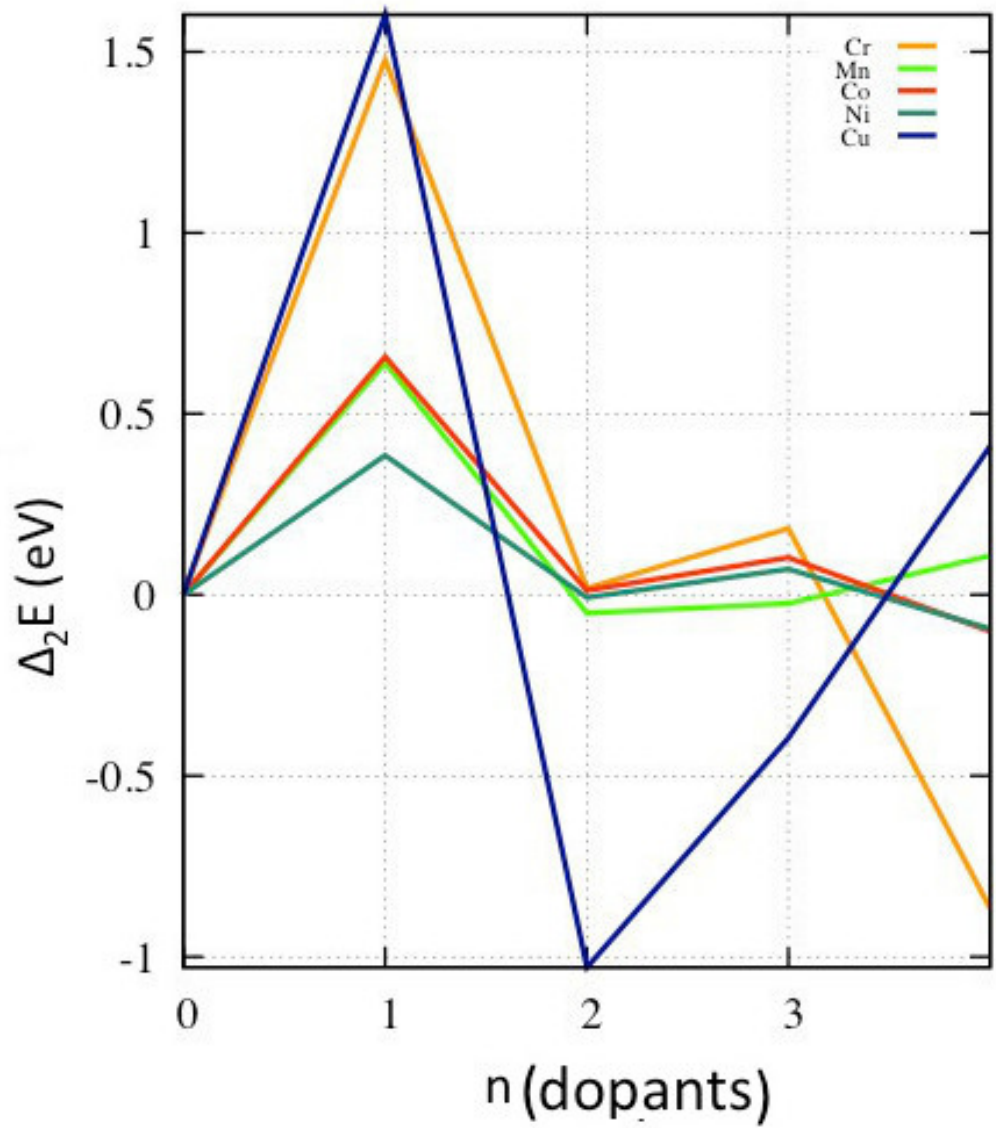

Fig. 4. The second energy difference, $\Delta_2 E$, versus n, the number of dopants

### 3.3 Density of States

For a metallic nanocluster of $Fe_{13}$, the density of states (DOS) in the vicinity of the Fermi level is important as it is responsible for chemical reactivity. To compare the effect of doping, we plotted first the pure $Fe_{13}$ spin polarized total DOS and the total projected DOS for d orbitals since it is the d electrons that vary due to doping by 3d transition elements.

As seen in Fig. 5, the majority channel of $Fe_{13}$ has many more states than the minority channel for states below the Fermi level and vice versa above the Fermi level. The d orbitals make up the majority of states except at the Fermi level for the majority channel and at higher energies above 2 eV. The effect of doping can be categorized into two classes as those of 1) Co, Ni and Cu, and 2) Cr and Mn. The DOS for $Fe_{13}$ doped with Co, Ni and Cu retain the general structure of the $Fe_{13}$ DOS. As the concentration is increased, the van Hove singularities have broadened, with the magnitude of the peaks decreasing for some or remaining constant for others. This indicates the change in the distribution of the DOS, the availability of more states with wider range of energies, not just those near the peaks. The changes in DOS are seen across the energy spectrum. It is worth noting, however, at or near the Fermi level in the majority channel, the change in DOS is very miniscule, only the size of the peak has slightly decreased monotonically, except for Cu at n = 4. By contrast, more states are formed in the minority channel in the vicinity of the Fermi level. In the second category, i.e. $Fe_{13}$ doped with Cr or Mn, the DOS have changed significantly. The DOS are quite different from that of the parent structure ($Fe_{13}$) without doping. In bulk, Cr and Mn are antiferromagnetic metals, while Fe, Co and Ni are ferromagnetic. It is not surprising then that the effect of Cr and Mn as dopants on the DOS of the nanocluster is quite different from that of Co and Ni. Overall, the surface states are created across the energy spectrum. At the Fermi level or the vicinity, these states are created in the minority channel as for Co, Ni and Cu. One interesting feature seen at the Fermi level is that for Mn and Cr, the states have decreased to miniscule level at n = 4, suggesting that doping with lower concentration may be preferable. Co and Cu also show the same trend of decreasing DOS at the Fermi level as the concentration increases. On the other hand, the Fermi level for Ni remains almost constant n = 1, 2, and 4, and shows a slight increase for n = 3.

Finally, the icosahedral symmetry of $Fe_{13}$ is broken very slightly due to very minor changes in bond lengths at the range of doping concentration considered in this work, therefore, we will not consider the Jahn – Teller effect. Our main interest in the DOS is to see how surface states are created as a result of doping and their availability for surface interaction to facilitate adsorption of carbon atoms for SWCNT growth or adhesion of SWCNT fragments to $Fe_{13}$ nanocatalyst.

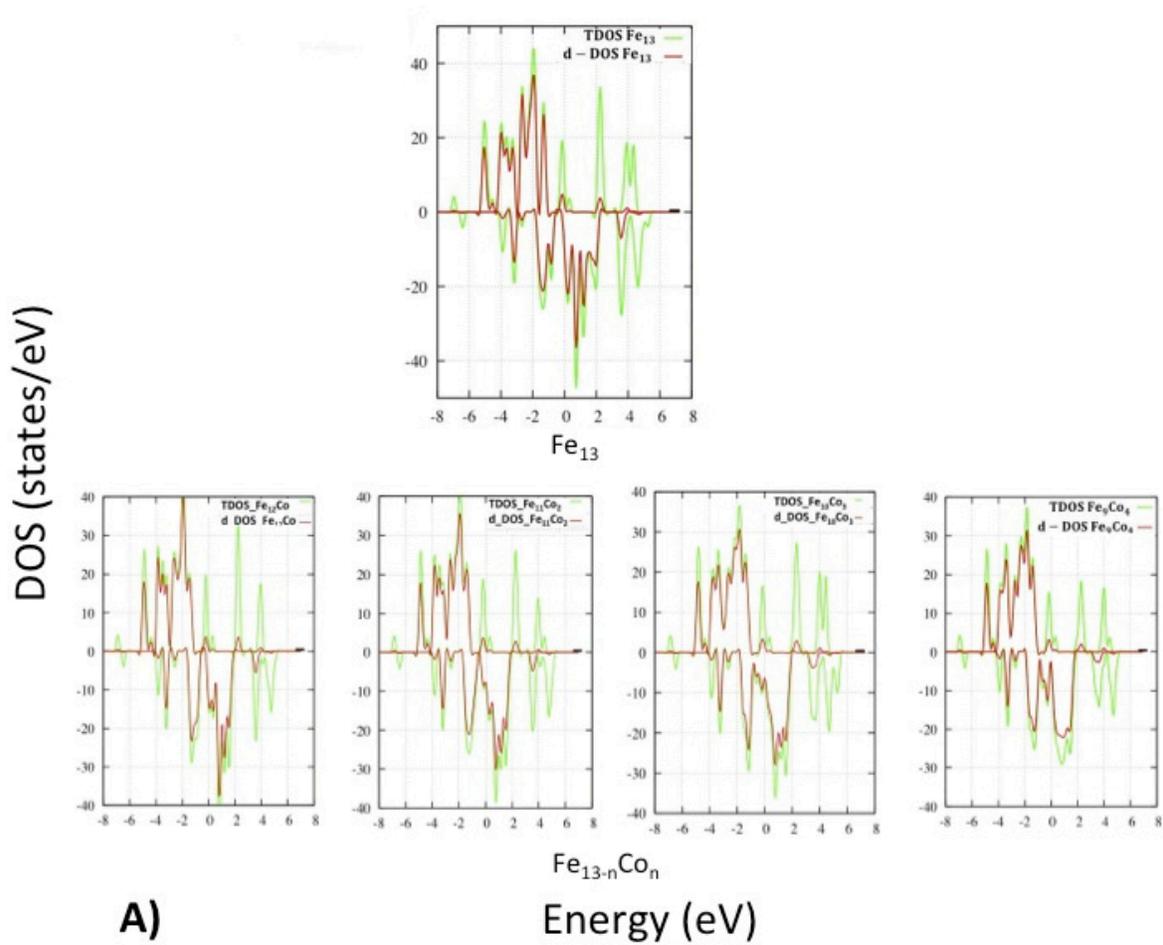

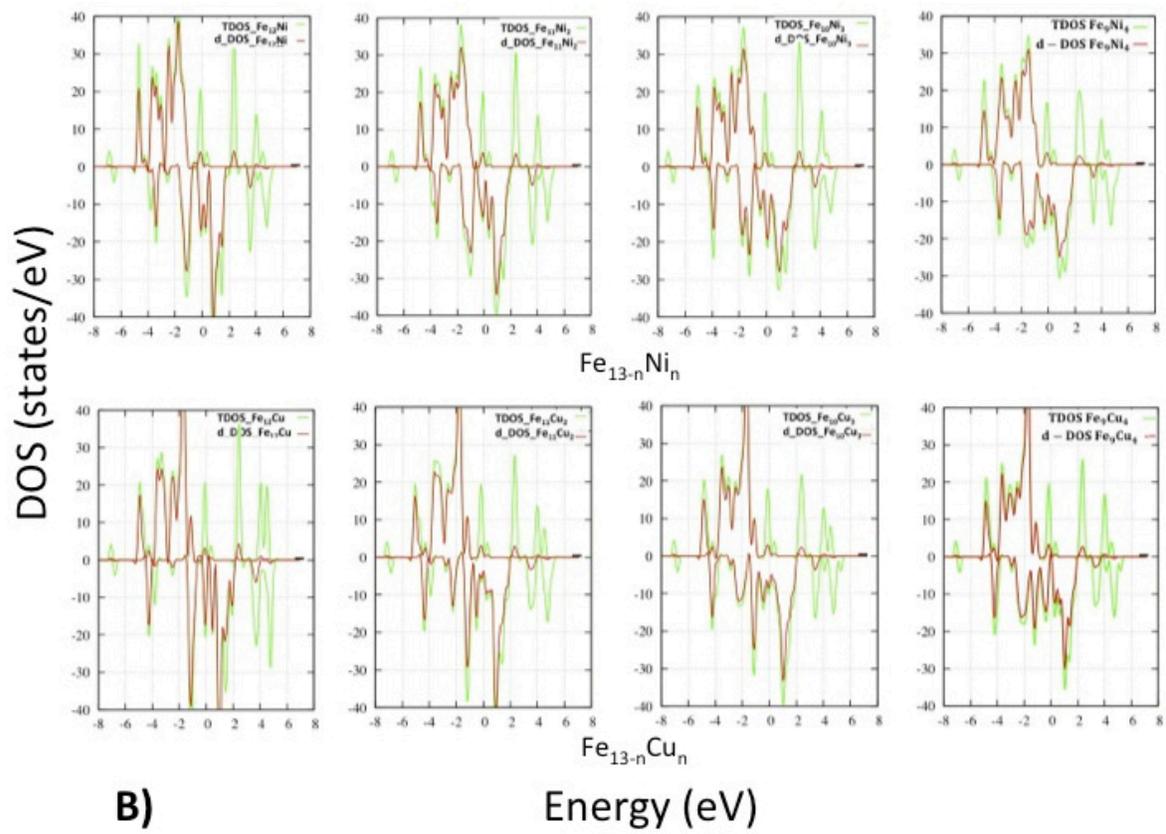

$Fe_{13-n}Ni_n$

$Fe_{13-n}Cu_n$

**B)**

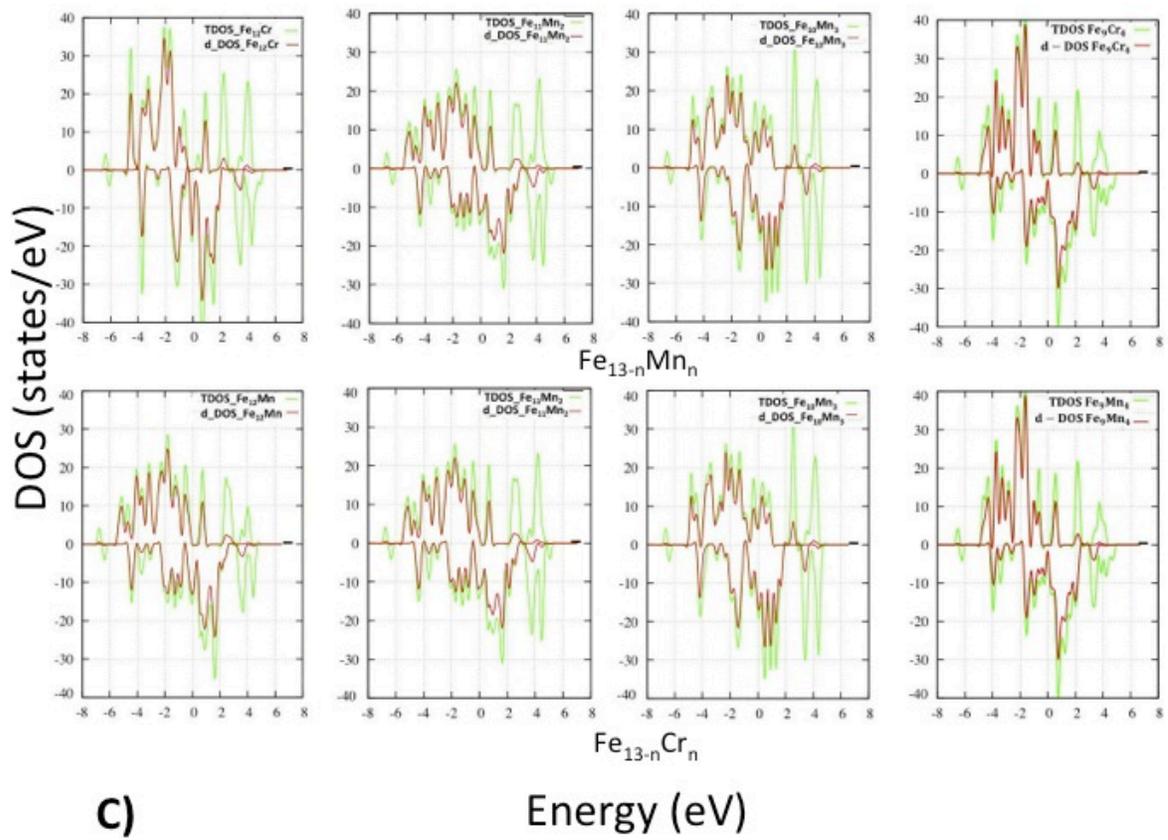

$Fe_{13-n}Mn_n$

$Fe_{13-n}Cr_n$

**C)**

FIG. 5. From top down, total DOS (green) and total d - projected DOS (brown) for A) $Fe_{13-n}Co_n$ and Fe, B) $Fe_{13-n}Cu_n$, $Fe_{13-n}Ni_n$, C) $Fe_{13-n}Cr_n$, $Fe_{13-n}Mn_n$. For each row, from left to right n = 1, 2, 3, and 4.

.

## 3. 4 Adsorption

We looked at the effect of doping on adsorption by placing a carbon atom on the triangular faces of the lowest energy structure of $Fe_9M_4$ where M = Cr, Mn, Co, Ni, and Cu. There are four possible adsorption sites: a triangle with all vertices consisting Fe atoms (P1), two vertices consisting of Fe and one of M (P2), one consisting of Fe and two of M (P3), and all of them consisting of M (P4).

| Element (M) | P1 | P2 | P3 | P4 |
| --- | --- | --- | --- | --- |
| Cr | -8.26 | -8.18 | -8.63 | NA |
| Mn | -8.04 | -8.10 | -8.19 | -8.26 |
| Co | -8.31 | -8.32 | -8.73 | NA |
| Ni | -8.45 | -8.56 | -8.52 | -8.53 |
| Cu | -8.40 | -8.29 | -7.65 | -7.08 |

Table 3. Adsorption energies in eV. NA stands for not applicable, i.e., the configuration does not exist.

The strongest binding energy for Cr resulted from P3, for Mn from P4, for Co from P3, for Ni from P2, and for Cu from P1. The C atom was adsorbed on the face for Cr, Mn, and Ni. In the case of Co (P3), the carbon atom ended in between the Co atoms, and in the case of Cu, the carbon atom led to bond breaking that resulted in an open structure. Our basic premise with regards to nucleation of carbon atoms[26] is that the carbon atoms first get adsorbed on the faces and then on bridge sites leading to a zigzag structure in the presence of ambient carbon atoms. Mn, Co and Ni also have face adsorptions for the remaining possibilities as well, so they are good candidates as a catalyst. Cr has face adsorptions, but the structures loose their icosahedral symmetry that is important in our model. Overall, Ni has the strongest bonding energy for all the possible configurations suggesting it would be the most easily doped element.

**3.5 Adhesion energy**

We looked at the trend of the adhesion energies between the (5, 0) SWCNT fragments consisting of 40 carbon atoms and and $Fe_{13-n}M_n$ where M=Cr, Mn, Co, Ni and Cu and n=4. The end of the SWCNT away from the nanocluster was passivated with hydrogen atoms. We also computed the adhesion energy, as given in equation 2, for pure $Fe_{13}$ cluster for comparison. We note that the adhesion energy is defined in such a way because we are only interested in the trend not the exact value. The passivation by H will add a constant value to overall energy and will not affect the trend. The total adhesion energies for (5, 0) are -37.5 eV for Cr, -37.3 eV for Mn, -37.9 eV for Co, -36.8 eV for Ni, -37.2 for Cu, and -38.5 eV for pure Fe. The relative total adhesion energy differences with respect to Fe are 1.0 eV, 1.2 eV, .6 eV, 1.7 eV, 1.3 eV, for Cr, Mn, Co, Ni and Cu respectively. For the 33% coverage of 3d elements, doping $Fe_{13}$ with Co results in the strongest adhesion energy and the weakest for Ni. However, the changes are within 1.0 ± 0.4 eV and are almost of the same magnitude, implying all of the doped structures can support SWCNT growth. It is worth noting also that the adhesion energy is maximum for pure $Fe_{13}$. To compare it with previous calculations, we subtracted the contribution of the hydrogen atoms[38], and divided the resulting energies by the number of bonds, which is 5 at the interface. We got the following adhesion energies: -5.51, -5.47, -5.58, -5.37 and -5.71 eV for M = Cr, Mn, Co, Ni, and pure Fe respectively. All the results above for Fe, Co and Ni are consistent with the findings of Larson et al.[39]

## 4. Conclusion

The icosahedral symmetry of $Fe_{13}$ must be retained in our model for chiral selectivity and anomalous cap formation.[26, 27] We found that doping with the lighter elements of Sc, Ti, and V leads to distortions of the icosahedral symmetry. Cr is on the borderline; one of its homotops is distorted. Icosahedral symmetry is retained for all homotops of Mn, Co, Ni and Cu. The stability analysis shows that the n = 1 is the most stable one. For n > 1, only Cu and Mn show the rise in stability after initial drop at n = 2. The DOS analysis shows two distinct features. Those for $Fe_{13}$ doped with Co, Ni, and Cu are very different from those doped with Cr and Mn. For all the dopants, the n = 1 doping has significant DOS of states around the Fermi level. The DOS at Fermi level decreases with concentration for Cr, Mn, Co and Cu and remains constant or shows slight increase for Ni. The adsorption analysis shows that Mn, Co and Ni have the lowest energies for adsorptions on the triangular faces without distorting the structure. Overall, Ni is the most easily adsorbed element with strong binding energies for all possible dopant configurations considered. Finally, the adhesion energy for SWCNT fragments is strongest for the undoped $Fe_{13}$. However, the adhesion energies for the doped $Fe_{13}$ are within 1.0 ± 0.4 eV of pure $Fe_{13}$ and can support SWCNT growth. Based on the overall results, we propose a core-shell type (even though there is only one atom at the core) structure with a central Mn atom surrounded by Fe surface shell atoms as the most favorable doped nanocatalyst for SWCNT nucleation. This is because doping the central atom with Mn leads to the most significant change in the DOS of the doped $Fe_{13}$. For doping with n > 1, the fact that the DOS for Ni remaining constant at the Fermi level or slightly increasing combined with its ease of adsorption with largest binding energies at all the possible configurations considered (except P3, Co binding energy is slightly larger) for the dopants makes it the best candidate for doping.

**ACKNOWLEDGEMENTS**
This research was supported by NSF grant: DMR-0804805. This research was also supported in part by the NSF through TeraGrid resources (the Stampede supercomputing system at TACC/UT Austin) provided by NCSA under grant number TG-DMR100055.